 \documentclass[preprint ]{aastex} 

\usepackage{emulateapj5}
\usepackage{apjfonts}

\makeatletter
\newenvironment{inlinefigure}{%
\def\@captype{figure}%
\noindent\begin{minipage}{0.999\linewidth}\begin{center}}
{\end{center}\end{minipage}\smallskip}
\makeatother

\shorttitle{Chandra LETG Spectrum of SS Cyg}
\shortauthors{Mauche}

 \slugcomment{Accepted for publication in {\it Astrophysical Journal\/} 2004 March    30}

\newcommand{\Mwd}{M_{\rm wd}}
\newcommand{\Rwd}{R_{\rm wd}}
\newcommand{\Msun}{\rm M_{\odot}}
\newcommand{\Mdot}{\dot{M}}
\newcommand{\lax}{{\lower0.75ex\hbox{ $<$ }\atop\raise0.5ex\hbox{ $\sim$ }}}
\newcommand{\gax}{{\lower0.75ex\hbox{ $>$ }\atop\raise0.5ex\hbox{ $\sim$ }}}

\begin{document}

\title{Chandra Low Energy Transmission Grating Spectrum of SS Cygni in
Outburst}

\author{Christopher W.\ Mauche\altaffilmark{1}}

\affil{Lawrence Livermore National Laboratory,
       L-473, 7000 East Avenue, Livermore, CA 94550}

\email{mauche@cygnus.llnl.gov}

\altaffiltext{1}{This paper is dedicated to the memory and accomplishments
of my colleague and friend Janet Aky\"uz Mattei, who died on 2004 March 22
after a long battle with acute myelogenous leukemia. Her passing is a great
loss to the astronomical community, both amateur and professional.}


\begin{abstract}
We have fitted the {\it Chandra\/} Low Energy Transmission Grating spectrum
of SS~Cygni in outburst with a single temperature blackbody suffering the
photoelectric opacity of a neutral column density and the scattering
opacity of an outflowing wind. We find that this simple model is capable of
reproducing the essential features of the observed spectrum with the blackbody
temperature $T_{\rm bl}\approx 250\pm 50$ kK, hydrogen column density
$N_{\rm H}\approx 5.0^{+2.9}_{-1.5}\times 10^{19}~\rm cm^{-2}$, fractional
emitting area $f\approx 5.6^{+60}_{-4.5}\times 10^{-3}$, boundary layer
luminosity $L_{\rm bl}\approx 5^{+18}_{-3}\times 10^{33}~\rm erg~s^{-1}$,
wind velocity $v\approx  2500~\rm km~s^{-1}$, wind mass-loss rate
$\Mdot_{\rm w}\approx 1.1\times 10^{16}~\rm g~s^{-1}$, and arbitrary values of
the wind ionization fractions of 20 ions of O, Ne, Mg, Si, S, and Fe. Given
that in outburst the accretion disk luminosity $L_{\rm disk}\approx 1\times
10^{35}~\rm erg~s^{-1}$, $L_{\rm bl}/L_{\rm disk}\approx 0.05^{+0.18}_{-0.03}$,
which can be explained if the white dwarf (or an equatorial belt thereon) is
rotating with an angular velocity $\Omega_{\rm wd}\approx 0.7^{+0.1}_{-0.2}$
Hz, hence $V_{\rm rot}\sin i\sim 2300~\rm km~s^{-1}$.
\end{abstract}

\keywords{accretion, accretion disks ---
          binaries: close ---
          novae, cataclysmic variables ---
          stars: individual (SS~Cygni) ---
          X-rays: binaries}


\section{Introduction}

According to simple theory, the boundary layer between the accretion disk and
the surface of the white dwarf is the dominant source of high energy radiation
in nonmagnetic cataclysmic variables (CVs). Unless the white dwarf is rotating
rapidly, the boundary layer luminosity $L_{\rm bl}\approx G\Mwd\Mdot /2\Rwd $,
where $\Mwd $ and $\Rwd $ are respectively the mass and radius of the white
dwarf and $\Mdot $ is the mass-accretion rate. When $\Mdot $ is low, as in
dwarf nova in quiescence, the boundary layer is optically thin to its own
radiation and its temperature is of order the virial temperature $T_{\rm vir}
= G\Mwd m_{\rm H}/3k\Rwd\sim 10$ keV. When $\Mdot $ is high, as in nova-like
variables and dwarf novae in outburst, the boundary layer is optically thick
to its own radiation and its temperature is of order the blackbody temperature
$T_{\rm bb} = (G\Mwd\Mdot /8\pi\sigma f\Rwd^3)^{1/4}\sim 100$ kK.

The compact nature of the boundary layer is established by the narrow
eclipses of the hard X-ray flux of high inclination dwarf novae in quiescence
\citep{woo95a, muk97, tes97, pra99a, whe03b} and the short periods [$P\sim
10~{\rm s}\sim 2\pi(\Rwd ^3/G\Mwd )^{1/2}$] of the oscillations of the soft
X-ray and extreme ultraviolet (EUV) flux of nova-like variables and dwarf
novae in outburst \citep{war04, mau04a}. However, an {\it extended\/} source
of soft X-ray and EUV flux is required to explain the lack of eclipses in
high inclination nova-like variables and dwarf novae in outburst \citep{nay88,
woo95b, pra99b, mau00, pra04}.

Using the {\it Extreme Ultraviolet Explorer\/} ({\it EUVE\/}) light curve and
spectrum of OY Car in superoutburst, \citet{mau00} argued that the source of
the extended EUV flux of high-$\Mdot $ nonmagnetic CVs is the accretion disk
wind. In this picture, in high inclination systems the boundary layer is
hidden by the accretion disk for all binary phases, but its flux is scattered
into the line of sight by resonance transitions of ions in the wind. Because
the scattering optical depths of these transitions are much higher than the
Thompson optical depth, the EUV spectra of such systems should be dominated
by broad lines.

Consistent with this picture, in outburst the EUV spectra of the high
inclination dwarf novae 
OY~Car ($i\approx 83^\circ $; \citealt{mau00}) and
WZ~Sge ($i\approx 75^\circ $; Wheatley \& Mauche 2004, in preparation)
are dominated by broad lines, while that of the low inclination dwarf novae
VW~Hyi ($i\sim    60^\circ $; \citealt{mau96}) and
SS~Cyg ($i\sim    40^\circ $; \citealt{mau95}) are dominated by the continuum.
 U~Gem ($i\approx 70^\circ $; \citealt{lon96}) is a transitional case, as its
EUV flux is partially eclipsed and its EUV spectrum contains of broad emission
lines superposed on a bright continuum.

The general case of the radiation transfer of the boundary layer flux through
the accretion disk wind is complicated, but schematically we expect that in
high inclination systems the EUV spectrum will be of the form $F_\lambda\,
e^{-\tau_\lambda ^\prime}\, (1-e^{-\tau_\lambda})$, while in low inclination
systems it will be of the form $F_\lambda\, e^{-\tau_\lambda ^\prime}
e^{-\tau_\lambda}$, where $F_\lambda$ is the intrinsic boundary layer
spectrum, $\tau_\lambda ^\prime$ is the photoelectric optical depth of the
wind and the interstellar medium, and $\tau_\lambda$ is the scattering
optical depth of the wind. \cite{mau00} showed that the second version of this
model describes the main features of the {\it EUVE\/} spectrum of OY~Car in
superoutburst, and we show here that the first version of this model describes
the main features of the {\it Chandra\/} Low Energy Transmission Grating (LETG)
spectrum of SS~Cyg in outburst. A preliminary discussion of this spectrum is
supplied by \citet{mau04a}, while the quasi-coherent oscillations of the EUV
flux is discussed by \citet{mau02}.

\section{Chandra Observation and Data Analysis}

Our pre-approved target-of-opportunity {\it Chandra\/} LETG plus High
Resolution Camera (HRC) observation of SS~Cyg was performed between 2001
January 16 $\rm 21^h13^m$ and January 17 $\rm 10^h50^m$ UT during the plateau
phase of a wide normal (asymmetric) dwarf nova outburst that began on January
12 (Fig.~1). When the observation began, SS~Cyg had been at maximum optical
light ($V\approx 8.5$) for approximately $3{1\over 2}$ days; long enough,
given the delay of approximately $1{1\over 2}$ days between the rise of the
optical and EUV light curves of SS~Cyg \citep{mau01}, for the EUV flux to
have reached maximum and for the system to have reached a quasi-steady state.
Indeed, the zero- and $\pm$ first-order LETG/HRC count rates were roughly
constant throughout the 47.3 ks observation at 1.5 and 4.9 $\rm counts~s^{-1}$,
respectively.

\begin{inlinefigure}
\centerline{\includegraphics{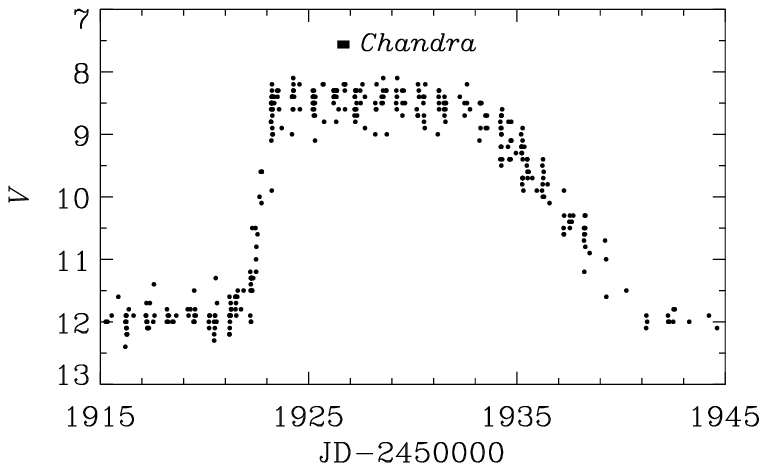}}
\figurenum{1}
\caption{AAVSO light curve of SS~Cyg \citep{mat03}. Black rectangle marks the
interval of the {\it Chandra\/} LETG observation.}
\end{inlinefigure}

The files used for analysis were created on 2002 January 7 by the pipeline
data reduction software using CIAO 2.0 with ASDCS version 6.5.1. Events were
extracted from the level 2 evt file, and the mean spectrum and source and
background region masks were extracted from the level 2 pha file. The
observation-average fluxed spectrum (units of $\rm erg~cm^{-2}~s^{-1}$ \AA
$^{-1}$) was derived from the $\pm $ first-order pha source and background
count spectra and the ``canned'' $\pm $ first-order LETG/HRC effective area
files distributed with {\it Chandra\/} CALBD 2.21. To avoid the HRC chip
gaps, the wave bands $\lambda =60.2$--67.2 \AA \ for the positive spectral
order and $\lambda =49.9$--57.2 \AA \ for the negative spectral order were
excluded. The spectrum used for subsequent analysis combines $\pm$ first
orders and is binned up in wavelength by a factor of four to $\Delta\lambda
=0.05$ \AA . 

As shown by \citet{mau04a}, the resulting LETG spectrum contains two distinct
components: (1) an X-ray component shortward of 42 \AA \ consisting of a
bremsstrahlung continuum and emission lines of H- and He-like C, N, O, Ne, Mg,
and Si and L-shell Fe (most prominently \ion{Fe}{17}) and (2) an EUV component
extending from 42 \AA \ to 130 \AA\ that consists of a continuum significantly
modified by a forest of broad absorption features (see Fig.~3). In the
72--130 \AA \ wave band, the LETG spectrum is essentially identical to that
measured by the {\it EUVE\/} short wavelength spectrometer during the 1993
August \citep{mau95} and 1996 October \citep{whe03a} outbursts of SS~Cyg. The
X-ray component of the LETG spectrum is not discussed here further, but is
understood to be due to the residual optically thin portion of the boundary
layer.

\section{Spectral Analysis}

To fit the EUV spectrum of SS~Cyg, we used the simple model described by
\citet{mau00}, which assumes that the boundary layer spectrum is that of a
single temperature blackbody suffering the photoelectric opacity of a neutral
column density and the scattering opacity of an outflowing wind. The boundary
layer spectrum is then specified by the blackbody temperature $T_{\rm bl}$,
fractional emitting area $f$ (hence luminosity $L_{\rm bl}=4\pi f\Rwd^2\sigma
T_{\rm bl}^4$), and hydrogen column density $N_{\rm H}$ (assuming the
\citealt{rum94} cross section with \ion{H}{1}, \ion{He}{1}, and \ion{He}{2}
with abundance ratios of 1:0.1:0.01). The properties of the wind are specified
by its mass-loss rate $\Mdot_0=4\pi r^2\mu m_{\rm H} nv=7.0\times 10^{15}~\rm
g~s^{-1}$ via fiducial values of the wind radius $r=10^{10}~\rm cm$, density
$n=10^{10}~\rm cm^{-3}$, and velocity $v=2500~\rm km~s^{-1}$, where the value
of the wind velocity was chosen to reproduce the observed blueshift of the
absorption lines. In the Sobolev approximation, the wind optical depth
$\tau_\lambda =(\pi e^2/m_{\rm e} c)\, n A X \lambda _{\rm ij} f_{\rm ij}\,
(g_{\rm i}/\Sigma g_{\rm i})\, (dv/dr)^{-1}$, where $dv/dr = 2500~\rm km~s^{-1}
/10^{10}~cm =0.025~s^{-1}$ is the wind velocity gradient, $\lambda _{\rm ij}$,
$f_{\rm ij}$, and $g_{\rm i}$ are respectively the wavelengths, oscillator
strengths, and statistical weights of the various resonance transitions
\citep{ver96}, $A$ are the elemental abundances relative to hydrogen
\citep{and89}, $X$ are the ionization fractions, and the other symbols have
their usual meanings. To account for the velocity of the wind along the line of
sight, the transition wavelengths were blueshifted by $\Delta\lambda _{\rm ij}
=\lambda _{\rm ij} v/c$. To match the observed widths of the absorption lines,
the opacity for all lines was distributed as a Gaussian with $\rm FWHM =
1.0$~\AA . Finally, we assumed that the source distance $d=160$ pc
\citep{har00} and the white dwarf mass $\Mwd = 1.0~\Msun $, hence the white
dwarf radius $\Rwd = 5.5\times 10^8$ cm.

With these assumptions, the free parameters of the model are the boundary
layer temperature $T_{\rm bl}$, fractional emitting area $f$, hydrogen column
density $N_{\rm H}$, wind mass-loss rate $\Mdot_{\rm w}=m\times \Mdot_0$, and
ionization fractions $X$. Operationally, the model was fit to the data between
45 \AA \ and 125 \AA \ using a grid search in the variables $T_{\rm bl}$,
$N_{\rm H}$, and $m$, with $X=1$ and $f$ determined by the model normalization
that minimizes $\chi^2$ between the model and the data. The resulting best-fit
model has $\chi^2$ per degree of freedom (dof) = $5961/1596
=3.74$ with $T_{\rm bl}=250$ kK, $N_{\rm H}=5.0\times 10^{19}~\rm cm^{-2}$,
$f=5.0\times 10^{-3}$, and $m=1.6$ and contains 34 lines of
\ion{O}{ 5}--\ion{O}{ 6},
\ion{Ne}{ 5}--\ion{Ne}{ 8},
\ion{Mg}{ 5}--\ion{Mg}{10},
\ion{Si}{ 5}--\ion{Si}{11},
\ion{S}{ 7}--\ion{S}{10}, 
\ion{Fe}{ 7}--\ion{Fe}{10}, and 
\ion{Fe}{12}--\ion{Fe}{15}
(31 ions of six elements) that depress the continuum by at least 10\%. Next,
$T_{\rm bl}$, $N_{\rm H}$, and $m$ were frozen and $X$ was allowed to vary for
each of these 31 ions. Table~1 shows the evolution of $\chi^2$ as sequential
steps were taken in this new parameter space. The resulting best-fit model has
$\chi^2/{\rm dof}=3333/1565=2.13$ with $f=5.6\times 10^{-3}$ and 41 lines of
\ion{O}{ 5}--\ion{O}{ 6},
\ion{Ne}{ 5}--\ion{Ne}{ 8},
\ion{Mg}{ 5}--\ion{Mg}{9},
\ion{Si}{ 6}--\ion{Si}{ 9},
\ion{S}{ 7},
\ion{Fe}{ 7}, \ion{Fe}{9}, \ion{Fe}{13}, and \ion{Fe}{15}
(20 ions of six elements) that depress the continuum by at least 10\%.
The resulting contours of $\chi^2$ and $f$ in the $[T_{\rm bl}, N_{\rm H}]$
parameter plane are shown in Figure~2 and the best-fit model is shown
superposed on the data in Figure~3.

With $\chi^2/{\rm dof}=2.13$, the fit shown in Figure~3 is unacceptable in
a statistical sense, but it clearly reproduces the essential features of the
observed spectrum. Ignoring the unmodeled \ion{C}{5} He$\alpha $ emission
feature at 41 \AA , the largest systematic residuals between the model and the
data are in the neighborhood of 48, 65, 86, and 92 \AA , where the data is
above the model, but this discrepancy probably could be accommodated in a more
detailed model by raising the continuum and increasing the wind optical depth.
The goodness of the fit is due in large part to the arbitrary variations of
the ionization fractions, which decrease $\chi^2/{\rm dof}$ from $5961/1596
=3.74$ to $3333/1565=2.13$ after 14 steps in $\chi^2$ space. The reasons for
the sequential improvements in the fit documented in Table~1 are often clear
[e.g., increasing $X$(\ion{Mg}{8}) increases the depth of the model absorption
features at $\approx 69.0$ \AA \ and $\approx 74.4$ \AA ; increasing
$X$(\ion{Si}{7}) increases the depth of the model absorption features at
$\approx 67.6$ \AA \ and $\approx 69.0$ \AA ; zeroing out $X$(\ion{Si}{5})
removes a strong model absorption feature at $\approx 96.4$ \AA ], but others
are less so [e.g., increasing $X$(\ion{Fe}{7}) increases the depth of a
number of absorption features longward of $\approx 102$ \AA ]. While the
individual values of the ionization fractions in Table~1 are not unreasonable,
as a {\it set\/} they make little sense physically.

\begin{center}
\begin{tabular}{lcc}
\multicolumn{3}{c}{TABLE 1}\\
\multicolumn{3}{c}{Model Ionization Fractions}\\
\hline
\hline
Ion& $X$& $\chi^2$\\
\hline
\hbox to 1.0in{\ion{Mg}{ 8}\leaders\hbox to 0.4em{\hss.\hss}\hfill}& 4.0& 5210\\
\hbox to 1.0in{\ion{Si}{ 7}\leaders\hbox to 0.4em{\hss.\hss}\hfill}& 3.2& 4742\\
\hbox to 1.0in{\ion{Fe}{ 7}\leaders\hbox to 0.4em{\hss.\hss}\hfill}& 10.& 4276\\
\hbox to 1.0in{\ion{Si}{ 5}\leaders\hbox to 0.4em{\hss.\hss}\hfill}& 0.0& 4089\\
\hbox to 1.0in{\ion{Mg}{ 7}\leaders\hbox to 0.4em{\hss.\hss}\hfill}& 1.6& 3938\\
\hbox to 1.0in{\ion{Fe}{13}\leaders\hbox to 0.4em{\hss.\hss}\hfill}& 4.0& 3775\\
\hbox to 1.0in{\ion{Mg}{ 5}\leaders\hbox to 0.4em{\hss.\hss}\hfill}& 2.5& 3660\\
\hbox to 1.0in{\ion{Mg}{ 6}\leaders\hbox to 0.4em{\hss.\hss}\hfill}& 0.5& 3567\\
\hbox to 1.0in{\ion{Fe}{12}\leaders\hbox to 0.4em{\hss.\hss}\hfill}& 0.0& 3511\\
\hbox to 1.0in{\ion{S}{  9}\leaders\hbox to 0.4em{\hss.\hss}\hfill}& 0.0& 3464\\
\hbox to 1.0in{\ion{Ne}{ 8}\leaders\hbox to 0.4em{\hss.\hss}\hfill}& 0.6& 3418\\
\hbox to 1.0in{\ion{Si}{10}\leaders\hbox to 0.4em{\hss.\hss}\hfill}& 0.0& 3383\\
\hbox to 1.0in{\ion{Mg}{ 8}\leaders\hbox to 0.4em{\hss.\hss}\hfill}& 3.2& 3352\\
\hbox to 1.0in{\ion{Si}{ 6}\leaders\hbox to 0.4em{\hss.\hss}\hfill}& 1.3& 3333\\
\hline
\end{tabular}
\end{center}

\begin{inlinefigure}
\centerline{\includegraphics{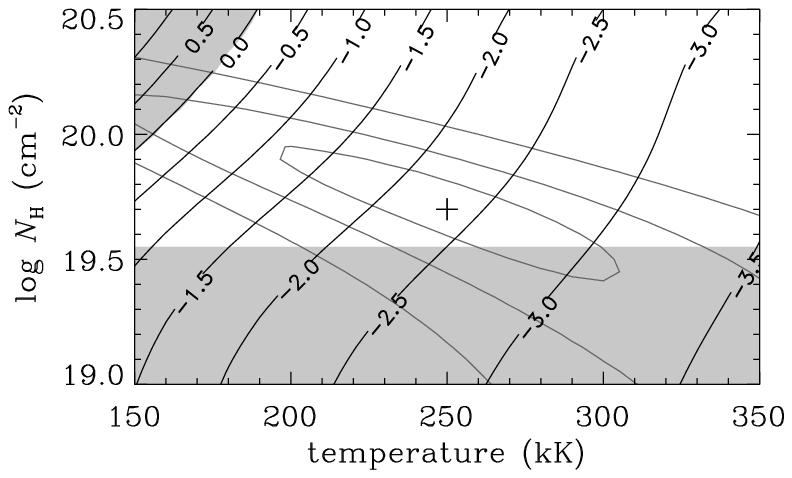}}
\figurenum{2}
\caption{Contours of $\chi^2=[1.5,3,6]\,\chi^2_{\rm min}$ ({\it gray curves\/})
and $\log f$ ({\it black curves\/}) in the $[T_{\rm bl}, N_{\rm H}]$ parameter
plane for the model described in \S 3. Regions of parameter space shaded gray
are excluded by the requirements that $N_{\rm H}\ge 3.5\times 10^{19}~\rm
cm^{-2}$ and $f\le 1$. Cross marks the parameters of the model shown in Fig.~3.}
\end{inlinefigure}

To see this, we used XSTAR v1.07 to calculated the ionization fractions of a
photoionized plasma for various values of the ionization parameter $\xi =
L_{\rm bl}/ nr^2$. With $L_{\rm bl}=4\pi f\Rwd^2\sigma T_{\rm bl}^4=4.7\times
10^{33}~\rm erg~s^{-1}$ and the fiducial values of $n$ and $r$, $\log\xi=3.7$.
Assuming that the plasma is optically thin and that the photoionizing source is
a 250 kK blackbody, the XSTAR runs demonstrate that the ionization parameter
must span the range $-1\lax\log\xi\lax 5$ to account for the range of ions
present in the model.

Given this result, we investigated whether it is possible to obtain good fits
to the data with less arbitrary sets of ionization fractions. In the first
attempt, we fixed the model parameters $T_{\rm bl}$ and $N_{\rm H}$ at their
nominal values, set the ionization fractions equal to their peak values for
$-1\le\log\xi\le 5$, and adjusted $m$ and $f$. This model produced only a small
improvement in the fit relative to that with $X=1$, since the peak ionization
fractions differ by factors of only a few. In the second attempt, we weighted
the ionization fraction distributions $X(\xi)$ by the function $e^{-\xi_{\rm
min}/\xi}e^{-\xi/\xi_{\rm max}}$, which suppresses the ionization fractions at
the lower and upper ends of the range. However, varying the exponential cutoffs
$\xi_{\rm min}$ and $\xi_{\rm max}$ and $m$ and $f$ again produced only a small
improvement in the fit relative to that with $X=1$. We conclude that our simple
model is capable of producing reasonably good fits to the data only if the
ionization fractions are allowed to vary arbitrarily.

\begin{figure*}
\figurenum{3}
\epsscale{2.03} 
\plotone{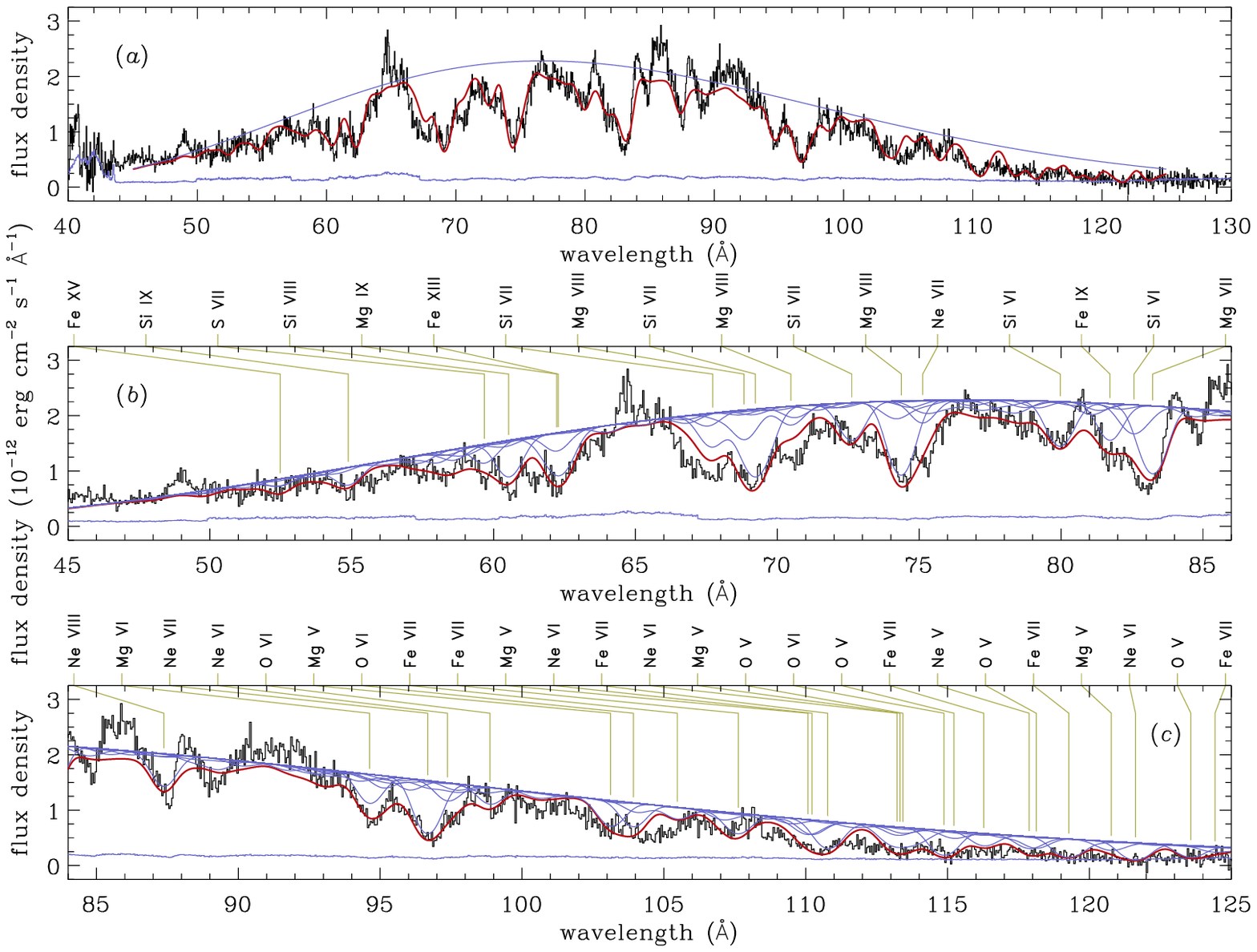}
\caption{{\it Chandra\/} LETG/HRC spectrum and best-fit model spectrum of SS
Cyg in outburst. Data are shown by the black histogram, the $1\,\sigma $ error
vector by the blue histogram, the absorbed blackbody continuum by the smooth
blue curve, and the net model spectrum by the thick red curve. In panels
({\it b\/}) and ({\it c\/}) the individual ion spectra are shown by the blue
curves and the strongest lines in the model are labeled.}
\end{figure*}

\section{Discussion}

We have fitted the {\it Chandra\/} LETG spectrum of SS~Cyg in outburst with a
single temperature blackbody suffering the photoelectric opacity of a neutral
column density and the scattering opacity of an outflowing wind. The best-fit
model, shown in Figure~3, has parameters $T_{\rm bl}=250$ kK, $N_{\rm H} = 
5.0\times 10^{19}~\rm cm^{-2}$, $f=5.6\times 10^{-3}$, $\Mdot_{\rm w}=1.1
\times 10^{16}~\rm g~s^{-1}$, and the values of $X$ listed in Table~1 ($X=1$
otherwise). Assuming that the wind is optically thin and photoionized by a
250 kK blackbody, the range of ions present in the model requires that the
ionization parameter $-1\lax\log\xi\lax 5$.

Given these results, in outburst the boundary layer luminosity $L_{\rm bl}
= 4.7\times 10^{33}~\rm erg~s^{-1}$. A lower limit on the value of this
quantity is imposed by the value of the interstellar H column density,
which \citet{mau88} estimated to be $3.5\times 10^{19}~\rm cm^{-2}$ based
on the curve-of-growth of interstellar absorption lines in high resolution
{\it International Ultraviolet Explorer\/} spectra of SS~Cyg in outburst. For
$-1\lax\log\xi\lax 5$, H and He are fully ionized, so the wind should have no
photoelectric opacity in the EUV bandpass, and $N_{\rm H}$ should be equal to
the interstellar H column density. With $N_{\rm H}=3.5\times 10^{19}~\rm
cm^{-2}$ and $T_{\rm bl}=300$ kK, Figure~2 shows that $f=1.1\times 10^{-3}$,
hence $L_{\rm bl}=1.7\times 10^{33}~\rm erg~s^{-1}$. At the other end of the
$\chi^2$ ellipse, with $T_{\rm bl}=200$ kK and $N_{\rm H}=7.9\times 10^{19}~\rm
cm^{-2}$, $f=6.6\times 10^{-2}$, hence $L_{\rm bl}= 2.3\times 10^{34}~\rm
erg~s^{-1}$. We conclude that a reasonable estimate of the boundary layer
luminosity $L_{\rm bl}\approx 5^{+18}_{-3}\times 10^{33}\, (d/160~{\rm
pc})^2~\rm erg~s^{-1}$.

How does this compare with the accretion disk luminosity? \citet{pol84} used
optical through far ultraviolet spectra of SS~Cyg in outburst to determine
that $L_{\rm disk}\approx 1\times 10^{35}\, (d/160~{\rm pc})^2~\rm erg~s^{-1}$,
so $L_{\rm bl}/L_{\rm disk}\approx 0.05^{+0.18}_{-0.03}$. Theoretically, we
expect that this ratio is equal to one unless the white dwarf is rotating
rapidly, in which case $L_{\rm bl}/L_{\rm disk}=(1-\omega)^2$, where $\omega
=\Omega_{\rm wd}/\Omega_{\rm K}(\Rwd )$, $\Omega_{\rm wd}=2\pi/P_{\rm spin}$
is the angular velocity of the white dwarf, and $\Omega_{\rm K}(\Rwd )
= (G\Mwd/\Rwd^3)^{1/2}\approx 0.9$ Hz is the Keplerian angular velocity of
material just above its surface \citep{pop95}. The white dwarf can be spun
up to the required rate by accreting an amount of mass $\Delta M=\Mwd - 
M_{\rm wd,i}$, where $M_{\rm wd,i}$ is the initial white dwarf mass, $\Mwd =
M_{\rm wd,i}\, (1-4k^2\omega/3)^{-3/4}$ is the final white dwarf mass, and
$k$ is the radius of gyration \citep{lan00}. With $k\approx 0.4$ and $\omega
= 1- (L_{\rm bl}/L_{\rm disk})^{1/2}\approx 0.78^{+0.08}_{-0.26}$, $\Delta
M\approx 0.13\, \Mwd\sim 0.1~\Msun $. If rapid rotation is the cause of
SS~Cyg's low boundary layer luminosity, the white dwarf angular velocity
$\Omega_{\rm wd} =\omega\Omega_{\rm K}(\Rwd )\approx 0.7^{+0.1}_{-0.2}$ Hz,
hence the spin period $P_{\rm spin}\approx 9^{+4}_{-1}$ s (comparable to the
period of the quasi-coherent oscillations), the rotation velocity $V_{\rm rot}
= \Omega_{\rm wd}\Rwd\approx 3800^{+400}_{-1300}~\rm km~s^{-1}$, and
$V_{\rm rot}\sin i\sim 2300~\rm km~s^{-1}$. This last quantity is rather
large, given that $V_{\rm rot}\sin i\approx 1200~\rm km~s^{-1}$ for WZ~Sge
and $V_{\rm rot}\sin i\lax 400~\rm km~s^{-1}$ for six other nonmagnetic CVs
\citep{sio99}.

Finally, we note that in the single scattering limit, conservation of momentum
limits the mass-loss rate of a radiatively driven wind to $\Mdot_{\rm max}
=L/V_\infty c$, where $L$ is the system luminosity and $V_\infty $ is the wind
terminal velocity. For SS~Cyg in outburst, $L=L_{\rm disk}+L_{\rm bl}\approx
1\times 10^{35}~\rm erg~s^{-1}$ and $V_\infty\approx 3000~\rm km~s^{-1}$, so
$\Mdot_{\rm max}\approx 1\times 10^{16}~\rm g~s^{-1}$. We derived a wind
mass-loss rate $\Mdot_{\rm w}\approx 1\times 10^{16}~\rm g~s^{-1}$ for SS~Cyg
in outburst, so $\Mdot_{\rm w}/\Mdot_{\rm  max}\approx 1$. In $\tau $~Sco
(B0 V), P~Cyg (B1 Ia$^+$), $\zeta $~Pup (O4 f), and $\epsilon $~Ori (B0 Ia)
$\Mdot_{\rm w}/\Mdot_{\rm max}=0.02$, 0.22, 0.33, and 0.65, but in WR1
(WN5) $\Mdot_{\rm w}/\Mdot_{\rm max}=60$ \citep{lam99}, so we cannot conclude
from this result that a mechanism other than radiation pressure is needed to
drive the wind of SS~Cyg.

\section{Summary}

We have fitted the {\it Chandra\/} LETG spectrum of SS~Cyg in outburst with a
single temperature blackbody suffering the photoelectric opacity of a neutral
column density and the scattering opacity of an outflowing wind. Figures 2 and
3 show that this simple model is capable of reproducing the essential features
of the observed spectrum with 
$T_{\rm bl}\approx 250\pm 50$ kK,
$N_{\rm H}\approx 5.0^{+2.9}_{-1.5}\times 10^{19}~\rm cm^{-2}$, 
$f\approx 5.6^{+60}_{-4.5}\times 10^{-3}$,
$L_{\rm bl}\approx 5^{+18}_{-3}\times 10^{33}~\rm erg~s^{-1}$, and the wind
velocity $v\approx 2500~\rm km~s^{-1}$, mass-loss rate 
$\Mdot_{\rm w}\approx 1.1\times 10^{16}~\rm g~s^{-1}$,
and ionization fractions $X$ listed in Table~1 ($X=1$ otherwise). Given that in
outburst the accretion disk luminosity $L_{\rm disk}\approx 1\times 10^{35}~\rm
erg~s^{-1}$, $L_{\rm bl}/L_{\rm disk}\approx 0.05^{+0.18}_{-0.03}$, which can
be explained if the white dwarf (or an equatorial belt thereon) is rotating
with an angular velocity $\Omega_{\rm wd}\approx 0.7^{+0.1}_{-0.2}$ Hz, hence
$V_{\rm rot}\sin i\sim 2300~\rm km~s^{-1}$.

We have now used our simple one dimensional spectral model to fit the {\it
EUVE\/} spectrum of OY~Car in superoutburst \citep{mau00} and the {\it
Chandra\/} LETG spectra of WZ~Sge in superoutburst (Wheatley \& Mauche 2004,
in preparation) and SS~Cyg in outburst. Although the model is unable to
reproduce the asymmetries of the EUV emission features of OY Car and WZ Sge,
or to produce an acceptable fit to the EUV spectrum of SS~Cyg even with
arbitrary values of the ionization fractions, we believe that it captures the
essential physics responsible for shaping the EUV spectra of these systems,
and that it returns reasonable estimates of the values of the physical
parameters that describe the boundary layer (effective temperature, emitting
area, luminosity) and wind (velocity and mass-loss rate) of high-$\Mdot $
CVs. To make further progress in this area, we need to expand the model from
one dimension to three dimensions; account for the velocity, density, and
ionization structure of the wind; and compute the paths of boundary layer
photons as they scatter in the wind. This could be accomplished with a Monte
Carlo radiation transfer code, which we are in the process of developing
\citep{mau04b}.

\acknowledgments

Our pre-approved {\it Chandra\/} target-of-opportunity observation of SS~Cyg
was made possible by the optical monitoring and alerts provided by the members,
staff (particularly E.\ Waagen), and director, J.\ Mattei, of the AAVSO, and by
the efforts of {\it Chandra\/} X-Ray Observatory Center Director H.\ Tananbaum,
Mission Planner K.\ Delain, and the {\it Chandra\/} Flight Operations Team at
MIT.
We acknowledge with thanks the variable star observations from the AAVSO
International Database contributed by observers worldwide and used in this
research.
Support for this work was provided by NASA through {\it Chandra\/} Award Number
GO1-2023A issued by the {\it Chandra\/} X-Ray Observatory Center, which is
operated by the Smithsonian Astrophysical Observatory for and on behalf of
NASA under contract NAS8-39073.
This work was performed under the auspices of the US Department of Energy by
University of California, Lawrence Livermore National Laboratory under contract
W-7405-Eng-48.



\begin{thebibliography}{}
\bibitem[Anders \& Grevesse(1989)]{and89}
         Anders, E., \& Grevesse, N. 1989, \gca , 53, 197
\bibitem[Harrison et al.(2000)]{har00}
         Harrison, T.~E., McNamara, B.~J., Szkody, P., \& Gilliland, R.~L.
         2000, \aj , 120, 2649 
\bibitem[Lamers \& Cassinelli(1999)]{lam99}
         Lamers, H.~J.~G.~L.~M., \& Cassinelli, J.~P. 1999, Introduction to
         Stellar Winds (Cambridge: Cambridge Univ.\ Press)
\bibitem[Langer et al.(2000)]{lan00}
         Langer, N., Deutschmann, A., Wellstein, S., \& H\"oflich, P.
         2000, \aap , 362, 1046 
\bibitem[Long et al.(1996)]{lon96}
         Long, K.~S., Mauche, C.~W., Raymond, J.~C., Szkody, P., \& 
         Mattei, J.~A. 1996, \apj , 469, 841
\bibitem[Mattei(2003)]{mat03}
         Mattei, J.~A. 2003, Observations from the AAVSO International
         Database, personal communication
\bibitem[Mauche(1996)]{mau96}
         Mauche, C.~W. 1996, in Cataclysmic Variables and Related Objects,
         ed.\ A.\ Evans \& J.~H.\ Wood (Dordrecht: Kluwer), 243
\bibitem[Mauche(2002)]{mau02}
         Mauche, C.~W. 2002, \apj , 580, 423 
\bibitem[Mauche(2004)]{mau04a}
         Mauche, C.~W. 2004, in X-Ray Timing 2003: Rossi and Beyond, ed.\
         P.\ Kaaret, F.\ K.\ Lamb, \& J.\ H.\ Swank (Melville, NY: AIP),
         in press [astro-ph/0401484]
\bibitem[Mauche et al.(2004)]{mau04b}
         Mauche, C.~W., Liedahl, D.~A., Mathiese, B.~F., Jimenez-Garate,
         M.~A., \& Raymond, J.~C. 2004, \apj , in press [astro-ph/0401328]
\bibitem[Mauche, Mattei, \& Bateson(2001)]{mau01}
         Mauche, C.~W., Mattei, J.~A., \& Bateson, F.~M. 2001, in Evolution
         of Binary and Multiple Stars, ed.\ Ph.~Podsiadlowski, S.~Rappaport,
         A.~R.\ King, F.~D'Antona, \& L.~Burderi (San Francisco: ASP), 367
\bibitem[Mauche \& Raymond(2000)]{mau00}
         Mauche, C.~W., \& Raymond, J.~C. 2000, \apj , 541, 924
\bibitem[Mauche, Raymond, \& C\'ordova(1988)]{mau88}
         Mauche, C.~W., Raymond, J.~C., \& C\'ordova, F.~A. 1988, \apj ,
         335, 829 
\bibitem[Mauche, Raymond, \& Mattei(1995)]{mau95}
         Mauche, C.~W., Raymond, J.~C., \& Mattei, J.~A. 1995, \apj , 446, 842
\bibitem[Mukai et al.(1997)]{muk97}
         Mukai, K., Wood, J.~H., Naylor, T., Schlegel, E.~M., \& Swank, J.~H. 
         1997, \apj , 475, 812
\bibitem[Naylor et al.(1988)]{nay88} 
         Naylor, T., et al. 1988, \mnras , 231, 237
\bibitem[Polidan \& Holberg(1984)]{pol84}
         Polidan, R.~S., \& Holberg, J.~B. 1984, \nat , 309, 528 
\bibitem[Popham \& Narayan(1995)]{pop95}
         Popham, R., \& Narayan, R. 1995, \apj , 442, 337 
\bibitem[Pratt et al.(1999a)]{pra99a}
         Pratt, G.~W., Hassall, B.~J.~M., Naylor, T., \& Wood, J.~H. 
         1999a, \mnras , 307, 413
\bibitem[Pratt et al.(1999b)]{pra99b}
         Pratt, G.~W., Hassall, B.~J.~M., Naylor, T., Wood, J.~H., \& 
         Patterson, J. 1999b, \mnras , 309, 847
\bibitem[Pratt et al.(2004)]{pra04}
         Pratt, G.~W., Mukai, K., Hassall, B.~J.~M., Naylor, T., \& Wood,
         J.~H. 2004, \mnras , 348, L49.
\bibitem[Rumph, Bowyer, \& Vennes(1994)]{rum94}
         Rumph, T., Bowyer, S., \& Vennes, S. 1994, \aj , 107, 2108
\bibitem[Sion(1999)]{sio99}
         Sion, E.~M. 1999, \pasp , 111, 532 
\bibitem[Warner(2004)]{war04}
         Warner, B. 2004, \pasp , 116, 115 
\bibitem[Wheatley, Mauche, \& Mattei(2003)]{whe03a}
         Wheatley, P.~J., Mauche, C.~W., \& Mattei, J.~A. 2003, \mnras ,
         345, 49
\bibitem[Wheatley \& West(2003)]{whe03b}
         Wheatley, P.~J., \& West, R.~G. 2003, \mnras , 345, 1009 
\bibitem[Wood et al.(1995)]{woo95a}
         Wood, J.~H., Naylor, T., Hassall, B.~J.~M., \& Ramseyer, T.~F.
         1995, \mnras , 273, 772
\bibitem[Wood, Naylor, \& Marsh(1995)]{woo95b}
         Wood, J.~H., Naylor, T., \& Marsh, T.~R. 1995, \mnras , 274, 31
\bibitem[van Teeseling(1997)]{tes97}
         van Teeseling, A. 1997, \aap , 319, L25
\bibitem[Verner, Verner, \& Ferland(1996)]{ver96}
         Verner, D.~A., Verner, E.~M., \& Ferland, G.~J.
         1996, Atomic Data and Nuc.\ Data Tables, 64, 1
\end{thebibliography}
\end{document}